\documentclass[aps,prl,superscriptaddress,twocolumn,twoside,a4paper,floatfix]{revtex4-1}
\usepackage{hyperref,graphicx,amsmath,amssymb,mathpazo,color}
\usepackage[normalem]{ulem}

\usepackage{enumitem}

\usepackage{pgfplots,tikz}

\usepackage[british]{babel}
\usepackage{amsfonts,epsfig}
\usepackage{hyperref}
\usepackage{latexsym}
\usepackage{epsfig}
\usepackage{amsmath}
\usepackage{amssymb}
\usepackage{amsfonts}
\usepackage{amsthm}
\usepackage{mathrsfs}
\usepackage{mathtools}
\usepackage{natbib}
\usepackage{color,verbatim,graphics}
\usepackage{psfrag}
\DeclareMathAlphabet{\mathrsfs}{U}{rsfs}{m}{n}
\DeclareMathAlphabet{\mathpzc}{OT1}{pzc}{m}{it}
\DeclareMathAlphabet{\matheus}{U}{eus}{m}{n}
\DeclareMathAlphabet{\mathbbold}{U}{bbold}{m}{n}

\newcommand{\CC}{\mathbb{C}}

\newcommand{\ba}{\begin{eqnarray}}
\newcommand{\be}{\begin{equation}}
\newcommand{\ee}{\end{equation}}

\newcommand{\ea}{\end{eqnarray}}
\newcommand{\ban}{\begin{eqnarray*}}
\newcommand{\ean}{\end{eqnarray*}}
\newcommand{\tr}{\operatorname{tr}}

\newcommand{\ket}[1]{|#1\rangle}
\newcommand{\bra}[1]{\langle#1|}

\newcommand{\ketbra}[2]{|#1\rangle\langle#2|}

\newcommand{\etal}{{\it{et al.}}}

\begin{document}

\title{Cyclic Einstein-Podolsky-Rosen Steering}

\author{Istv\'an M\'arton}
\email{marton.istvan@atomki.hu}
\affiliation{MTA Atomki Lend\"ulet Quantum Correlations Research Group, Institute for Nuclear Research, P.~O. Box 51, H-4001 Debrecen, Hungary}

\author{S\'andor Nagy}
\email{sandor.nagy@science.unideb.hu}
\affiliation{Department of Theoretical Physics, University of Debrecen, P.~O. Box 5, H-4010 Debrecen, Hungary}

\author{Erika Bene}
\email{bene@atomki.hu}
\affiliation{MTA Atomki Lend\"ulet Quantum Correlations Research Group, Institute for Nuclear Research, P.~O. Box 51, H-4001 Debrecen, Hungary}  

\author{Tam\'as V\'ertesi}
\email{tvertesi@atomki.hu}
\affiliation{MTA Atomki Lend\"ulet Quantum Correlations Research Group, Institute for Nuclear Research, P.~O. Box 51, H-4001 Debrecen, Hungary}  

\begin{abstract}
Einstein-Podolsky-Rosen (EPR) steering is a form of quantum correlation that exhibits a fundamental asymmetry in the properties of quantum systems. Given two observers, Alice and Bob, it is known to exist bipartite entangled states which are one-way steerable in the sense that Alice can steer Bob's state, but Bob cannot steer Alice's state. Here we generalize this phenomenon to three parties and find a cyclic property of tripartite EPR steering. In particular, we identify a three-qubit state whose reduced bipartite states are one-way steerable for arbitrary projective measurements. Moreover, the three-qubit state has a cyclic steering property in the sense, that by arranging the system in a triangular configuration the neighboring parties can steer each others' states only in the same (e.g. clockwise) direction. That is, Alice can steer Bob's state, Bob can steer Charlie's state, and Charlie can steer Alice's state, but not the other way around. 
\end{abstract}

\maketitle

Quantum entanglement is a remarkable phenomenon without counterpart in classical physics~\cite{horo_review,gt_review}. Notably, it gives rise to nonlocal correlations between distant particles as it was pointed out by Einstein, Podolsky and Rosen (EPR)~\cite{EPR}. Later Bell~\cite{bell} proved that nonlocality is inherent to quantum theory. Today, Bell nonlocality is considered a fundamental feature of the theory and plays an important role in quantum information processing~\cite{horo_review,bell_review,vcbook}.

The concept of steering (also known as EPR steering) was proposed by Schr\"odinger in 1935~\cite{schrodinger}, which concept brought novel insight into the study of nonlocal correlations~\cite{pauldani_review,uola_review}. Consider two distant observers -- say, Alice and Bob -- who share a pair of two spin-($1/2$) particles in the maximally entangled singlet state 
\begin{equation}
\ket{\psi_-}=(\ket{0}_A\ket{1}_B-\ket{1}_A\ket{0}_B)/\sqrt 2.
\label{singlet}
\end{equation}
Alice can steer the state of Bob's system by performing a measurement on her system. In particular, if Alice projects by measuring her share of the state into the state  
\begin{equation}
\ket{v_A}=a\ket{0} +  b\ket{1},
\label{vA}
\end{equation}
Bob’s  system  immediately  collapses  to the orthogonal state 
\begin{equation}
\ket{v_B}=-b^*\ket{0} +  a^*\ket{1},
\label{vB}
\end{equation}
where  $*$  means  complex  conjugation. Note that due to normalization $aa^*+bb^*=1$, the coefficients $a$ and $b$  of Bob's state can be given explicitly by two angles $\theta$ and $\varphi$:
\begin{align*}
a=&\cos(\theta/2),\\
b=&\sin(\theta/2)\exp(i\varphi),
\end{align*}
where $a$ is real valued, since a global phase of the state~(\ref{vA}) is unobservable. The two angles $\theta$ and $\phi$ can be adjusted to arbitrary values by Alice by performing a well-chosen measurement on her system. Hence Alice can prepare different states for Bob, that is, she can steer Bob's state.

Originally, EPR steering was studied in the context of continuous variable systems \cite{reid,reid_review}, however, the effect was soon formalized by Wiseman~\etal~\cite{wiseman} as a quantum information task for general multipartite systems. EPR steering can also be seen as a form of quantum correlation that is intermediate between entanglement and Bell nonlocality~\cite{wiseman,saunders}. To illustrate these properties, let us consider the two-qubit singlet state~(\ref{singlet}). By adding some white noise to it we obtain the one-parameter family of two-qubit Werner states~\cite{werner}
\begin{equation} 
\rho_W(p) = p\ket{\psi_-}\bra{\psi_-}+(1-p)\mathbb{I}_4/4,
\label{rhow}
\end{equation}
where $(1-p)\in\left[0,1\right]$ is a noise parameter. One can now ask about the critical limit $p$ above which the state~(\ref{rhow}) is entangled, EPR steerable, and Bell nonlocal for arbitrary projective measurements. We list below the three different cases.
\begin{enumerate}[label=(\roman*)]
\item The state~(\ref{rhow}) is entangled if and only if $p>1/3$. This can be directly seen by using Peres' positivity of partial transpose (PPT) criterion~\cite{peres}.
\item  The state~(\ref{rhow}) is EPR steerable if and only if $p>1/2$ as shown by Werner~\cite{werner} and Wiseman~\etal~\cite{wiseman}. 
\item The state~(\ref{rhow}) is Bell nonlocal for $p>0.6964$~\cite{peter} and admits a local hidden variable model for $p\le 0.6829$ ~\cite{hirsch}. These bounds arise from a connection to the Grothendieck constant of order three~\cite{AGT}.
\end{enumerate}
The above non-overlapping bounds show that entanglement, steering and Bell nonlocality are different under projective measurements. However, the more general case of POVM measurements was also considered in Ref.~\cite{marco15} and the same conclusion has been drawn in this case as well. 

Steering finds applications in quantum information tasks, such as quantum key distribution~\cite{cyril,KWW}, randomness generation~\cite{law,paul18,curchod}, and channel discrimination~\cite{piani}. More recently, it has also been linked to quantum metrology~\cite{yadin}.

Experimental investigations have been reported~\cite{wittman,bennet,smith}, in particular the steerability property of the family of two-qubit states~(\ref{rhow}) has been analyzed in detail~\cite{saunders}. In addition, steering has been used as a tool for detecting entanglement in Bose-Einstein condensates~\cite{he1,fadel,kunkel} and atomic ensembles~\cite{he2}. Notably, in 2012, a loophole-free EPR steering experiment has been performed~\cite{wittman} (see also Ref.~\cite{weston}). We note that the more recent loophole-free Bell experiments~\cite{hensen} also demonstrate loophole-free EPR steerability. 

A distinctive feature of EPR steering is the asymmetry between the role of observers~\cite{wiseman,bowles}. In particular, this asymmetry is not present in the phenomenon of entanglement and Bell nonlocality. A steering test can be understood as the task of distributing entanglement from an untrusted Alice to a trusted Bob, a task formalized in Ref.~\cite{wiseman}. Concretely, consider two particles in different locations, which are controlled by Alice and Bob. Alice tries to convince Bob that they share an entangled state $\rho_{AB}$ of these two particles. Bob, however, does not trust Alice, and therefore asks her to steer the state of his particle using different measurements. Suppose that Alice can choose to perform $m$ different measurements labeled by $x=(1,\ldots,m)$ on her particle. Denote the POVM elements of her  outcome $a$ for a given setting $x$ by $M_{a|x}$. These POVM elements satisfy $M_{a|x}\ge 0$ for each choice of $x$ and outcome $a$, and we also have $\sum_a M_{a|x}=\mathbb{I}_d$, where $d$ is the dimension of the Hilbert space of Alice's subsystem. We will focus primarily on two-qubit states $\rho_{AB}\in\CC^2\otimes\CC^2$ shared by Alice and Bob, and on non-degenerate projective measurements $M_{a|x}$ for Alice, which can be written in the following form:
\begin{align*}
M_{0|x}=&\ket{v_A}\bra{v_A},\\
M_{1|x}=&\mathbb{I}_2-\ket{v_A}\bra{v_A},
\end{align*}
where $\ket{v_A}$ has the form~(\ref{vA}). The set of conditional states $\{\rho_{a|x}\}$ that Alice can prepare for Bob by measuring $x$ and obtaining $a$ forms the so-called steering assemblage. This set is given by the formula
\begin{equation}
\rho_{a|x}=\tr_A{\left(M_{a|x}\otimes \mathbb{I} \rho_{AB}\right)},
\label{assemb}
\end{equation}
where 
\begin{equation}
p(a|x)=\tr{(\rho_{a|x})}
\label{pax}
\end{equation}
is the probability that Alice obtains outcome $a$ for her setting $x$. Note that the states~(\ref{assemb}) are subnormalized in general, that is, $\tr{(\rho_{a|x})}<1$, however, $\sum_{a}{\tr{(\rho_{a|x})}}=1$ holds true for all $x$. In particular, the assemblage~(\ref{assemb}) carries all the information about the EPR steering setup.

We say that a state $\rho_{AB}$ demonstrates steering from Alice to Bob (or put differently, Alice can steer the state of Bob) if Bob's assemblage~(\ref{assemb}) cannot be written in the so-called local hidden state (LHS) form~\cite{wiseman}:
\begin{equation} 
\rho_{a|x} = \sum_\lambda  \omega(\lambda) p_\lambda(a|x) \rho_\lambda, 
\label{LHS}
\end{equation}
where $\lambda$ represents a classical random variable known to Alice with an arbitrary probability distribution $\omega(\lambda)$. That is, we have $\sum_{\lambda}\omega(\lambda)=1$ and $\omega(\lambda)\ge 0$ for all $\lambda$. Note that (\ref{LHS}) defines a so-called local hidden state strategy. In fact, the assemblage~(\ref{LHS}) can be prepared by Alice and Bob without sharing an entangled quantum state: shared randomness (characterized by the variable $\lambda$ and Bob's qubit states $\rho_{\lambda}$) is sufficient to produce the assemblage~(\ref{LHS}).
 
Hence, in the steering problem, Bob's task is to determine whether the states $\rho_{a|x}$ in the assemblage~(\ref{assemb}) admit a decomposition of the form~(\ref{LHS}). If this is the case, then Bob will not be convinced that entanglement is present. Conversely, if it can be shown that the assemblage~(\ref{assemb}) cannot be written in the form~(\ref{LHS}), then this indicates the presence of entanglement, and we say that the state $\rho_{AB}$ is steerable. This steerability can be  conveniently proved using the so-called EPR steering inequalities (which we discuss in our first scenario).

Let us also remark that a decomposition of the form~(\ref{LHS}) for a given finite number of settings $m$ does not imply that the underlying state $\rho_{AB}$ is unsteerable. It can well be the case that as the number of settings $m$ increases, the assemblage~(\ref{assemb}) can be no longer written in the LHS form~(\ref{LHS}) and thus becomes steerable. Hence, more generally, we say that a state $\rho_{AB}$ is unsteerable from Alice to Bob if the assemblage $\{\rho_{a|x}\}$ in (\ref{assemb}) admits a decomposition of the form (\ref{LHS}) for all possible measurements $M_{a|x}$. That is, Alice in general has to consider an infinite number of measurement settings $x$. We will mainly focus on projective measurements, in which case we say that the state $\rho_{AB}$ is unsteerable from Alice to Bob for projective measurements. 

We also note that the above definition of EPR steering treats the roles of the two observers differently. The question, already raised in Ref.~\cite{wiseman}, is whether there exists a bipartite entangled state $\rho_{AB}$ such that Alice can steer Bob's state, but Bob cannot steer Alice's state. This phenomenon, which has been called one-way steering, was first investigated theoretically in continuous variable systems for a restricted class of measurements~\cite{olsen1,olsen2}. Then a simple class of one-way steerable two-qubit states was found for projective measurements~\cite{bowles}. This phenomenon was further studied in other two-qubit systems~\cite{paul,bowles16,nguyen}. Finally, the problem has been settled down by finding a one-way steerable two-party state for the most general POVM measurements~\cite{marco15}. On the experimental side, early examples for one-way steering were presented for continuous variable systems involving Gaussian measurements \cite{handchen}. On the other hand, for discrete systems one-way steering including the general case of POVMs, was first demonstrated experimentally by Wollmann~\etal~\cite{wollmann}. 

Let us mention that the study of steering is not restricted to the bipartite case. Indeed, multipartite steering can be viewed as a semi-device-independent task~\cite{erik11}, where some of the parties are trusted and some of them are untrusted. In this scenario, for instance genuine tripartite entanglement~\cite{horo_review,gt_review,szilard} can be detected through the phenomenon of EPR steering~\cite{daniel15}. Furthermore, monogamy relations have also been studied in the context of steering for three-qubit systems~\cite{milne}.  

\begin{figure}[t!]
\includegraphics[trim=0 220 180 40,clip,width=10cm]{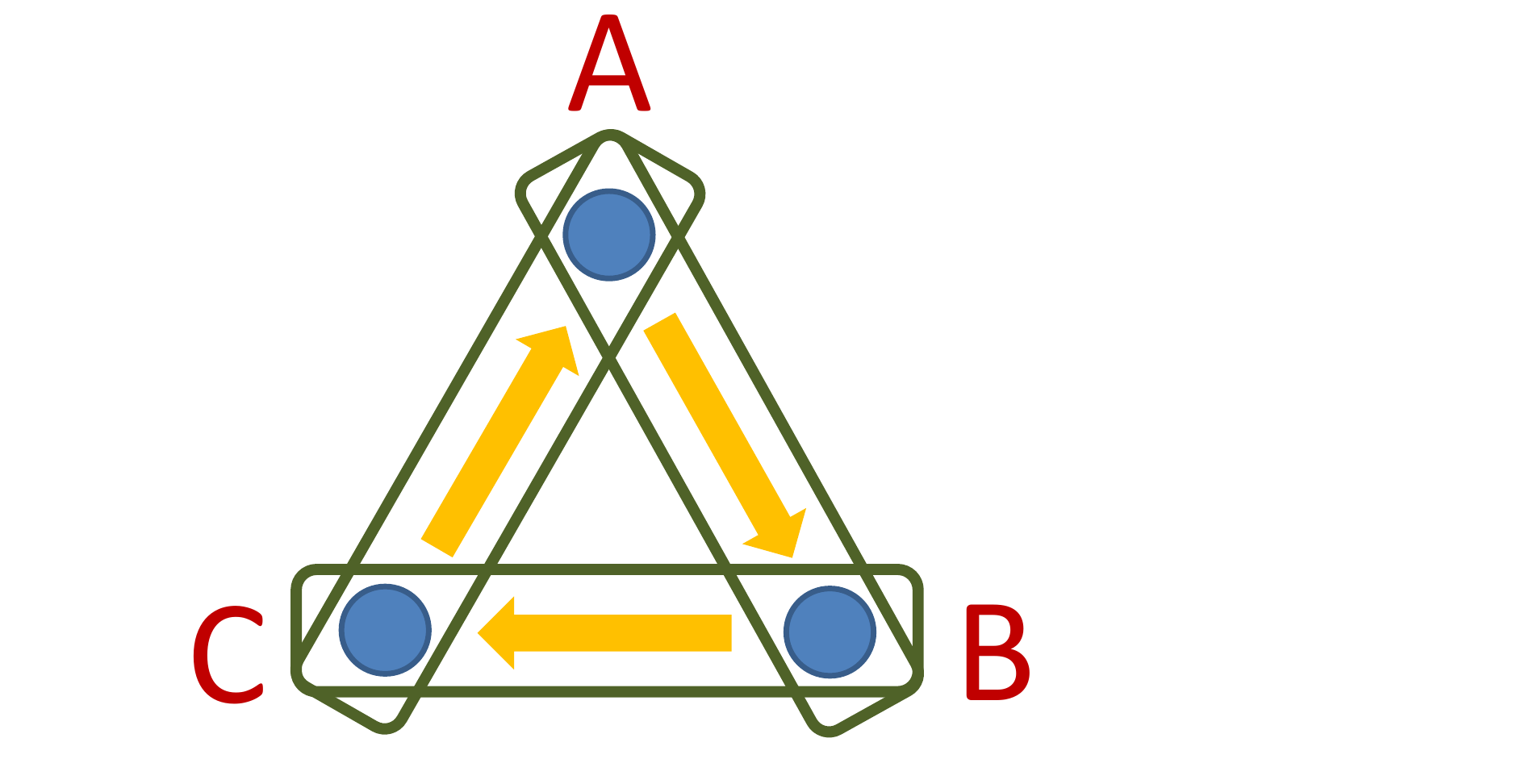}
\caption{Setup for cyclic EPR steering. The figure depicts a three-qubit state with its two-party reduces states in green rectangles. The yellow arrows indicate that steering takes place in the clockwise direction. That is, steering occurs from $A$ to $B$, from $B$ to $C$, and from $C$ to $A$. However steering is not possible in the other (anticlockwise) direction, even if the untrusted party can perform arbitrary projective measurements.} \label{fig_ABC}
\end{figure}

\emph{Our main result.}---We present a class of three-qubit entangled states which exhibits a cyclic steering property of quantum correlations. Specifically, we consider a three-qubit state in a triangular configuration shared by three partners Alice, Bob and Charlie, which state has the following property (see also Fig.~\ref{fig_ABC}): If any qubit is removed from the tripartite system, the remaining two-qubit state is one-way steerable. Let us denote by $P_1$ and $P_2$ the two parties left, and by $\rho_{P_1P_2}$ their respective reduced two-qubit state. Let us use the notation $P_1\to P_2$ if the state $\rho_{P_1P_2}$ demonstrates steering from $P_1$ to $P_2$, but it does not demonstrate steering from $P_2$ to $P_1$ considering arbitrary projective measurements (the case of POVM measurements and higher dimensional systems are briefly discussed in Appendix~D). 

In our particular case we prove the following steering properties of the $A-B-C$ tripartite system (see Fig.~\ref{fig_ABC}): $A\to B$, $B\to C$ and $C\to A$. To this end, we consider a translationally invariant three-qubit state $\rho_{ABC}$, that is, we have $\rho_{ABC}=S\rho_{ABC}$, where $S$ is the right-shift operator: 
\begin{equation}
S=\sum_{i,j,k=0}^1\ket{ijk}\bra{jki}.
\label{rightS}
\end{equation}
This state has two-qubit reduced states with the following property: 
\begin{equation}
\rho_{AB}=\rho_{BC}=\rho_{CA},
\label{ABdir}
\end{equation}
where $\rho_{AB}=\tr_C(\rho_{ABC})$ denotes the two-qubit reduced state of Alice and Bob. If we swap Alice and Bob, 
\begin{equation}
\rho_{BA}=V\rho_{AB}V^{\dagger},
\label{rhoBA}
\end{equation}
where $V$ is the two-qubit flip operator, we do not usually have $\rho_{AB}=\rho_{BA}$. However, this property holds true for permutationally invariant three-qubit states $\rho_{ABC}$, such as the famous W~\cite{Wstate} and GHZ states~\cite{GHZ}.

With the above symmetry of the state, $\rho_{ABC}=S\rho_{ABC}$, our task then boils down to find a translationally invariant three-qubit state $\rho_{ABC}$ for which the two-qubit reduced state $\rho_{AB}$ is one-way steerable. That is, the state $\rho_{AB}$ does not demonstrate steering from Bob to Alice for any set of projective measurements of Bob. However, the state $\rho_{AB}$ is steerable from Alice to Bob, i.e., it does violate a specific steering inequality with well-chosen measurements of Alice. The proof of the existence of such a state $\rho_{ABC}$ will be based on the recent geometrical approach of Nguyen~\etal~\cite{nguyen}. 

The structure of the paper is as follows. Our starting point is the construction of a family of translationally invariant three-qubit states $\rho_{ABC}$. With these states in hand, we solve the above stated problem in two steps. (i) As a preliminary step, we first find a state from this family for which the two-qubit marginal $\rho_{AB}$ does violate a six-setting steering inequality. However, the swapped two-qubit state $\rho_{BA}$ does not violate the same steering inequality considering arbitrary projective measurements. This result already proves an asymmetric property of the two-qubit marginal $\rho_{AB}$. Although, it does not solve the problem yet we originally formulated. (ii) We present a translationally invariant three-qubit state $\rho_{ABC}$ for which the reduced states $\rho_{AB}=\rho_{BC}=\rho_{CA}$ are one-way steerable for projective measurements. Crucial to the demonstration of these properties is a class of three-qubit states, the construction of which is discussed below.

\emph{The state.}---We consider a scenario featuring three remote parties, Alice, Bob and Charlie, who share the following state: 
\begin{equation}
\rho(p) = \frac{p}{3}\sum_{i=1}^3 \ket{\psi_i}\bra{\psi_i}+(1-p)\mathbb{I}_8/8,
\label{rho3q}
\end{equation}
where $\ket{\psi_1}$ is a generic pure three-qubit state 
\begin{equation}
\ket{\psi_1}=\sum_{i,j,k=0}^1 c_{ijk}\ket{i}_A\ket{j}_B\ket{k}_C,
\label{psi1cijk}
\end{equation}
and the other two states, $\ket{\psi_2}$ and $\ket{\psi_3}$, are related to $\ket{\psi_1}$ as follows:
\begin{equation}
\ket{\psi_{(n+1)\pmod{3}}}=S^n\ket{\psi_1} 
\end{equation}
for $n\ge 1$, where $S$ is the right-shift operator defined by (\ref{rightS}). Observe the cyclic property $\ket{\psi_{1}}=S^3\ket{\psi_1}$ and the translationally invariance of the three-qubit state~(\ref{rho3q}). We also emphasize that $\rho(p)$ is completely defined by the pure state $\ket{\psi_1}$ and the noise parameter $(1-p)$. With the above definition of the state, we then proceed to our first scenario. 

\emph{Scenario 1.} 
Here we present a bipartite steering inequality and show its violation using the reduced two-qubit marginals of the state~(\ref{rho3q}) and specific measurements $M_{a|x}$ for Alice. Note that Alice's $\pm 1$ observables are defined by $A_x=M_{0|x}-M_{1|x}$. Our steering inequality takes the following form:
\begin{equation}
\sum_{x=1}^m\sum_{a=0}^1\tr(F_{a|x}\rho_{a|x})\le L, 
\label{SI}
\end{equation}
where $L$ is the maximum for the left-hand-side functional to be obtained with an assemblage of the LHS form~(\ref{LHS}). Violation of this inequality proves that the steering assemblage~(\ref{assemb}) cannot be reproduced by an LHS model~(\ref{LHS}). Let us use the following functional due to Saunders~\etal~\cite{saunders} for (\ref{SI}):
\begin{equation}
F_{a|x}=(-1)^a B_x,
\label{SIsaunders} 
\end{equation}
where the traceless $2\times 2$ Hermitian matrices $B_x$ are $B_x=\vec b_x\cdot\vec\sigma$, where $\vec b_x$, $x=(1,\ldots,m)$ are some unit vectors and $\vec\sigma=(\sigma_x,\sigma_y,\sigma_z)$ denotes the vector of Pauli matrices. The maximum value $L$ can be obtained by solving the following integer programming problem
\begin{equation}
L=\max_{a_x=\pm 1}\left[\lambda_{\rm max}\left(\sum_{x=1}^m{a_x B_x}\right)\right]=
\max_{a_x=\pm 1}\left\|\sum_{x=1}^m{a_x\vec b_x}\right\|,
\label{Lmax}
\end{equation}
where $\|\vec v\|$ stands for the Euclidean norm of vector $\vec v$, and $\lambda_{\rm max}(X)$ denotes the largest eigenvalue of $X$. So, we have to consider a total of $2^m$ strings $[a_1,a_2,\ldots,a_m]$ (where $a_x=\pm 1, x=1,\ldots,m$) for Alice to obtain $L$. For small $m$ (e.g., for $m<20$), this task can be solved on a desktop computer by an exhaustive enumeration of all possible strings.

On the other hand, we use the formula~(\ref{assemb}) for $\rho_{a|x}$ to compute the maximum quantum value of the left-hand-side expression in (\ref{SI}) for fixed $\rho_{AB}$ and $B_x$ matrices. We then use $A_x=M_{0|x}-M_{1|x}$ to obtain 
\begin{equation}
Q(\rho_{AB})\equiv\max_{-\mathbb{I}\le A_x\le \mathbb{I}}\sum_{x=1}^m{\tr(A_x\otimes B_x \rho_{AB})}.
\label{qmax}
\end{equation}
The above optimization task~(\ref{qmax}) can be carried out for each $x$ separately. Indeed, let us write
\begin{equation} 
\tr(A_x\otimes B_x \rho_{AB})=\tr(A_x G_x), 
\label{AxGx}
\end{equation}
where we defined
\begin{equation}
G_x=\tr_B(\mathbb{I}\otimes B_x\rho_{AB}), 
\label{Gx}
\end{equation}
which results in
\begin{equation}
Q=\max_{-\mathbb{I}\le A_x\le \mathbb{I}}\tr(A_xG_x)=\tr\sqrt{G_xG_x^{\dagger}},
\label{qmax2}
\end{equation}
i.e., the maximum is given by the trace norm of $G_x$, where the optimal observables of Alice are 
\begin{equation}
A_x=\sum_{i=1,2}\text{sign}(\lambda_{x,i})\ketbra{v_{x,i}}{v_{x,i}},
\label{Axopt}
\end{equation}
where the eigendecomposition of $G_x$  is given as $G_x=\sum_{i=1,2}\lambda_{x,i}\ketbra{v_{x,i}}{v_{x,i}}$.

Let us now present our results for the steering inequality~(\ref{SI}) using the functional~(\ref{SIsaunders}). Namely, we choose $B_x=\vec b_x\cdot\vec\sigma$, in which case $B_x$ can be interpreted as Bob's observables in (\ref{qmax}). In particular, we consider measurement settings for Bob based on the regular icosahedron that have 12 vertices with six antipodal pairs $\pm \vec b_x$, $x=1,\ldots,6$. The endpoints of the vectors $\vec b_x$ are marked in blue in figure~\ref{fig_blochs}. The coordinates of the vertices are given by the formulas~(\ref{vecbx}) in Appendix~A.

\begin{figure}[t!]
\includegraphics[trim=140 270 40 280,clip,width=10cm]{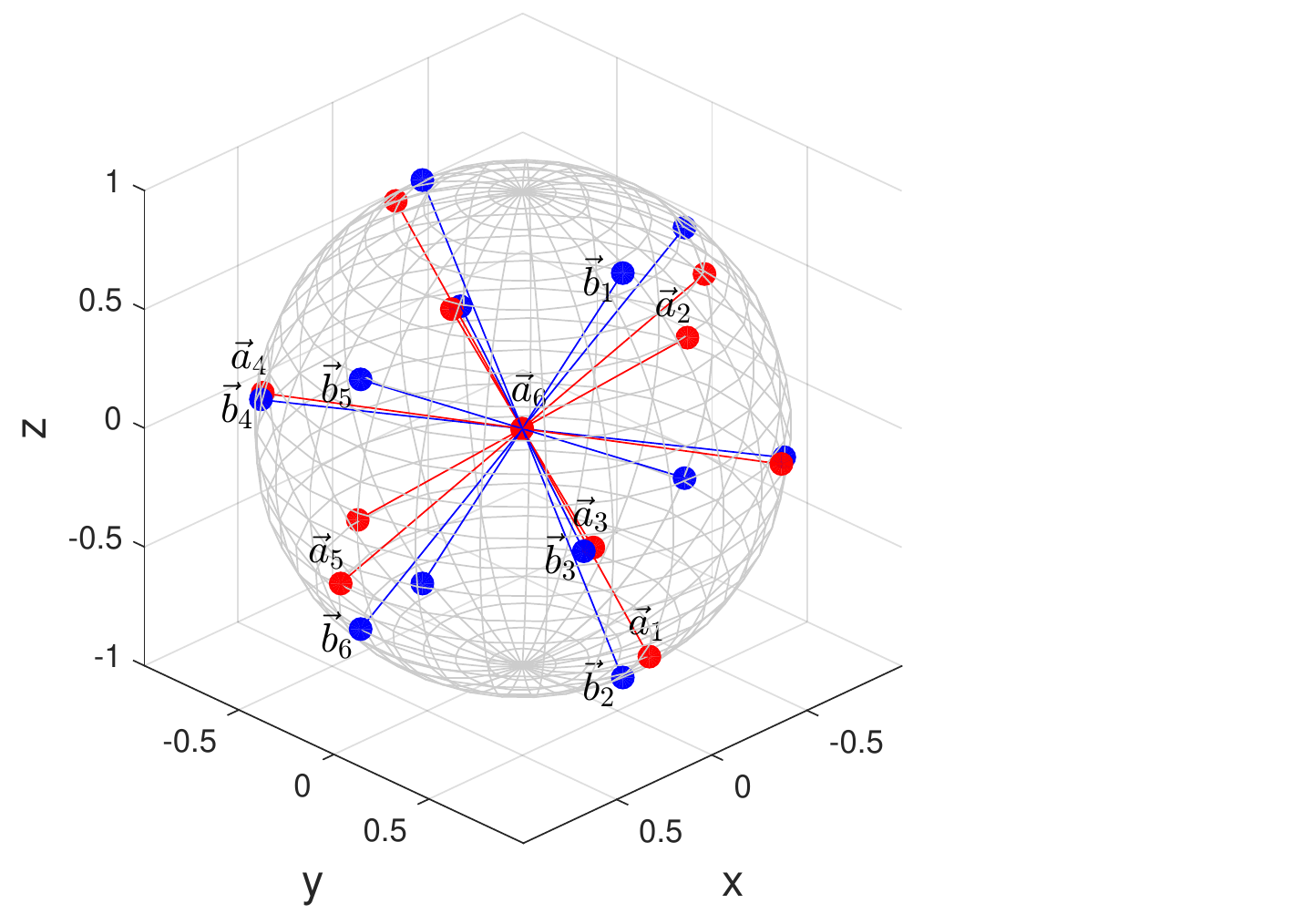}
\caption{Plot for the Bloch vectors of Alice and Bob measurements in the EPR steering scenario 1. Each measurement axis is defined by a pair of circles. The red circles correspond to Alice's Bloch vectors $\pm\vec a_x$, the blue circles to Bob's Bloch vectors $\pm\vec b_x$. Note that $\vec a_6$ is the zero-length vector corresponding to a degenerate measurement.} \label{fig_blochs}
\end{figure} 

Defining Bob's observables $B_x=\vec b_x\cdot\vec\sigma$ by the six icosahedral unit vectors $\vec b_x$, we get the following maximum for $L$ in (\ref{Lmax}):
\begin{align*}
L=\left\|\sum_{x=1}^6 a_x \vec b_x\right\|=\left\|v/n\right\|=1+\sqrt 5\simeq 3.2361,
\end{align*} 
where the vector $v=[2+2\phi,0,-2\phi]$. Alice's corresponding strategies are $a_1=-1$ and $a_x=1$ for $x=(2,\ldots,6)$. 

Our goal now is to find the reduced state $\rho_{AB}$ of the translationally invariant state~(\ref{rho3q}) such that: 
\begin{align}
&Q(\rho_{AB})>L,\nonumber\\
&Q(\rho_{BA})\le L,
\label{Qvalid}
\end{align}
where $\rho_{AB}=\tr_C{\rho(p)}$, with $\rho(p)$ given by (\ref{rho3q}), and the swapped state $\rho_{BA}$ is given by~(\ref{rhoBA}). Note that for a permutationally invariant (PI) state $\rho_{ABC}$, we have $\rho_{AB}=\rho_{BA}$ and both conditions in~(\ref{Qvalid}) cannot be fulfilled simultaneously. Therefore, we need to guarantee that $\rho(p)$ is not PI. We will perform a heuristic search in order to find a feasible $\rho_{AB}$ in~(\ref{Qvalid}). To do so, we transform the feasibility problem~(\ref{Qvalid}) into a constrained optimization task:
\begin{align}
\max &\quad Q(\rho_{AB})\nonumber\\ 
s.t.&\quad Q(\rho_{BA})\le L,
\label{COtask}
\end{align}
where we set $p=1$ in Eq.~(\ref{rho3q}). Hence, in the above task the only optimization variables are the $\alpha_{ijk}$ coefficients of $\ket{\psi_1}$ in (\ref{psi1cijk}). In addition we set $\alpha_{111}=0$ and restrict the search to real-valued coefficients $\alpha_{ijk}$. One variable is spared due to the normalization condition $\sum_{i,j,k}\alpha_{ijk}^2=1$, leaving $2^3-2=6$ optimization variables. 

To solve the above task we use the Nelder-Mead simplex algorithm~\cite{NM} by replacing the problem with an unconstrained problem, where we subtract a penalty term from the objective function $Q(\rho_{AB})$. Specifically, we maximize the following objective value
\begin{equation}
q=Q(\rho_{AB})-c\max\left[0,L-Q(\rho_{BA})\right]
\label{Qobj}
\end{equation}
over the $\alpha_{ijk}$ coefficients of $\ket{\psi_1}$. We set $c=2$ in the course of optimization, and $\max\left[x_1,x_2\right]$ gives the larger value between $x_1$ and $x_2$. The algorithm was run 500 times and a couple of times we found the following (presumably) best objective $q=3.2688$ where $Q(\rho_{AB})=3.2688$ and $Q(\rho_{BA})=3.2360$, which values satisfy the conditions~(\ref{Qvalid}). The corresponding (unnormalized) state is 
\begin{align}
\ket{\psi_1} = & +0.069455\ket{000} + \ket{001} +\ket{010}\nonumber\\
&-0.762707\ket{011} + 0.604546\ket{100}\nonumber\\
&-0.475110\ket{101} -0.762707\ket{110}.  
\label{optpsi1}
\end{align}
The negativity --- an entanglement measure~\cite{vidal} --- of the bipartite state $\rho_{AB}$ defined by $\ket{\psi_1}$ in (\ref{optpsi1}) and $p=1$ in Eq.~(\ref{rho3q}) is given by $E_N=0.1517$. In Appendix~A we provide Alice's optimal observables $A_x=\vec a_x\cdot\vec\sigma$ for this scenario, corresponding to the maximum value $Q(\rho_{AB})=3.2688$ in Eq.~(\ref{qmax}). 
In particular, $\ket{\psi_1}$ in (\ref{optpsi1}) through $\rho_{AB}$ defines $G_x$ in (\ref{Gx}) which in turn provides the $A_x=\vec a_x\cdot\vec\sigma$ observables~(\ref{Axopt}). The corresponding Bloch vectors $\vec a_x$ are given by (\ref{vecax}) in Appendix~A, which are represented in Fig.~\ref{fig_blochs}. The figure shows the icosahedron configuration of Bob's unit vectors $\vec b_x$ (blue markers) and Alice's six Bloch vectors $\vec a_x$ (red markers), $x=1,\ldots,6$. We next turn to our main problem of finding a genuine cyclic EPR-steerable state for projective measurements.
 
\emph{Scenario~2.}---We are looking for a three-qubit state in the form~(\ref{rho3q}), where each two-qubit marginal $\rho_{AB}=\rho_{BC}=\rho_{CA}$ is one-way steerable considering arbitrary projective measurements. For this purpose, we will rely on the recent two-qubit steerability result of Nguyen~\etal~\cite{nguyen}. In that work the critical radius $R(\sigma_{AB})$ of an arbitrary two-qubit state $\sigma_{AB}$ is defined and it is shown that $\sigma_{AB}$ is unsteerable from Alice to Bob if and only if $R(\sigma_{AB})\ge 1$. The critical radius has an operational meaning as well, $1-R(\sigma_{AB})$ measures the distance from the state $\sigma_{AB}$ to the border of the unsteerable states relative to the separable state $(\mathbb{I}_A/2)\otimes\sigma_B$, where $\sigma_B=\tr_A(\sigma_{AB})$ is Bob's reduced state. Moreover, Ref.~\cite{nguyen} gives an efficient numerical method to compute the critical radius for any two-qubit state considering arbitrary projective measurements. Namely, given a two-qubit state $\sigma_{AB}$, the numerical algorithm in Ref.~\cite{nguyen} gives both an upper bound $R^{\rm out}$ and a lower bound $R^{\rm in}$ on the critical radius $R$ of the state:
\begin{equation}
R^{\rm in}(\sigma_{AB})\le R(\sigma_{AB})\le R^{\rm out}(\sigma_{AB}).
\label{Rgap}
\end{equation}
The size of the interval above is related to a parameter $N$ of the method of Ref.~\cite{nguyen}. This parameter defines the number of vertices $N$ of an inscribed polyhedron of the unit sphere, and in the implemented version the largest parameter is $N=252$, which gives a small gap (typically smaller than $0.01R^{\rm in}$) between the upper and lower bound values in~(\ref{Rgap}).

Our goal is then to find a one-way steerable state $\rho_{AB}$ for projective measurements using the numerical procedure in Ref.~\cite{nguyen}, where $\rho_{AB}=\tr_C{\rho(p)}$, where $\rho(p)$ defines the three-qubit state~(\ref{rho3q}). We set $p=1$, and assume the most general form of the state $\ket{\psi_1}$ in Eq.~(\ref{psi1cijk}). 
If we happen to find that for some $\ket{\psi_1}$ both $R(\rho_{AB})<1$ and $R(\rho_{BA})\ge1$ are satisfied, then we have found a one-way steerable state $\rho_{AB}$, and we have achieved our goal. Formally, our task reduces to the following feasibility problem. Find a three-qubit state $\rho(p=1)$ in the form of Eq.~(\ref{rho3q}) with two-qubit reduced state $\rho_{AB}$ such that
\begin{align}
R_1&\equiv R^{\rm out}(\rho_{AB})<1,\nonumber\\
R_2&\equiv R^{\rm in}(\rho_{BA})\ge 1.
\label{R1R2}
\end{align}
Such inequalities entail the existence of $R(\rho_{AB})<1$ and $R(\rho_{BA})\ge 1$, thus directly proving that $\rho_{AB}$ is one-way steerable and eventually proving the existence of cyclic EPR steering for a state in Eq.~(\ref{rho3q}). To find such a state, similarly to our steering scenario~1, we run the Nelder-Mead heuristic search. This time we maximize the objective value
\begin{align}
q=&R_2-R_1-c_1\max\left[0,R_1-1\right]-c_2\max\left[0,1-R_2\right]\nonumber\\
&-c_3(\mathcal{H}(R_1-1)+\mathcal{H}(1-R_2))
\label{maxq}
\end{align}
over all coefficients of $\ket{\psi_1}$, where we set $c_i$ positive, such as $c_1=c_2=2c_3=1$, where $\mathcal{H}$ above denotes the Heaviside function. By setting $c_1\neq c_2$ we can search for asymmetric solutions, that is when the average value $(R_1+R_2)/2$ tends to differ from 1. However, the problem with this simple scheme is that the running time for a single evaluation of $q$ in (\ref{maxq}) by the parameter $N=252$ takes about an hour on a standard desktop computer. Therefore, the Nelder-Mead algorithm becomes very time consuming. Note that in the complex-valued case we have to optimize over $2\times2^3-1 = 15$ free variables defining the three-qubit pure state $\ket{\psi_1}$, which usually requires thousands of iterations to achieve convergence. To increase efficiency, we may choose to optimize (\ref{maxq}) with a smaller parameter, say $N=92$, for a couple of iteration steps and then switch to the more time consuming $N=252$ case. In practice, we have chosen an even simpler (i.e., less time consuming) solution for this preprocessing step. Namely, we used the following analytical upper bound value of $R(\sigma_{AB})$ for an arbitrary two-qubit state $\sigma_{AB}$~\cite{nguyen}:
\begin{equation}
R(\sigma_{AB})\le 2\pi N_T|\det(T)|\equiv R_a(\sigma_{AB}),
\end{equation}
where both $N_T$ and $\det(T)$ are analytical functions of $\sigma_{AB}$ with its closed form expression given by Ref.~\cite{jevtic}. In this case, we ran an unconstrained Nelder-Mead search to maximize the expression 
\begin{equation}
q=R_a(\rho_{BA})-R_a(\rho_{AB})-c\max\left[0,R_a(\rho_{BA})-\delta\right]
\label{maxq2}
\end{equation}
over $\rho_{AB}$, where the optimization variables are the coefficients of $\ket{\psi_1}$ in (\ref{psi1cijk}). Note that the last subtracted (positive) term $\delta$ penalizes the objective function whenever $R_a(\rho_{BA})>\delta$. We chose the threshold $\delta=1.2$ because of our empirical investigations, where we found that $R_a(\rho_{BA})$ can be considerably larger than $R^{\rm out}(\rho_{BA})$ in (\ref{Rgap}) for the parameter $N=252$. The optimization~(\ref{maxq2}) provides us with a state $\ket{\psi_1}$, which we pass to the further Nelder-Mead search~(\ref{maxq}) with parameter $N=252$. As a result of this two-step optimization procedure, we managed to obtain true one-way steerable states $\rho_{AB}$. We found the simplest state $\ket{\psi_1}$ defined by 7 real parameters (by setting $c_2>c_1$ and $c_3=1$), where the state vector $\ket{\psi_1}$ (up to normalization) looks as follows:
\begin{align}
\ket{\psi_1} = & \ket{000} -0.321193\ket{001} -0.477021\ket{010}\nonumber\\
& +0.045221\ket{011} -0.718592\ket{100}\nonumber\\
& + 0.213715\ket{101} -0.0482\ket{110}.      
\label{optpsi1simp}
\end{align}
The lower and upper bounds on the critical radius of the state $\rho_{AB}$ and its swapped state $\rho_{BA}$ are
\begin{align}
R^{\rm in}(\rho_{AB})&=0.9925396,\nonumber\\
R^{\rm out}(\rho_{AB})&=0.99822006,\nonumber\\
R^{\rm in}(\rho_{BA})&=1.0000028,\nonumber\\
R^{\rm out}(\rho_{BA})&=1.0065179,
\label{RinRout}
\end{align}
Since $R^{\rm out}(\rho_{AB})<1<R^{\rm in}(\rho_{BA})$ above holds, our state $\rho(p=1)$ with $\ket{\psi_1}$ in (\ref{optpsi1simp}) indeed defines a feasible solution to the phenomenon of cyclic EPR steering. Also by direct computation, the entanglement negativity of $\rho_{AB}$ is given by $0.0630$, which is to be compared to the negativity $1/2$ of the maximally entangled singlet state~(\ref{singlet}). Let us define the gap
\begin{equation}
\Delta(\rho_{AB})=R^{\rm in }(\rho_{BA})-R^{\rm out}(\rho_{AB})
\label{delta}
\end{equation}
between the feasible limits $R^{\rm in }(\rho_{BA})\ge1$ and $R^{\rm out}(\rho_{AB})<1$.
Note that the above example $\rho(p=1)$ with $\ket{\psi_1}$ in (\ref{optpsi1simp}) does not provide the largest possible gap. In fact, we have found states with larger gaps $\Delta(\rho_{AB})$ among generic real-valued states $\ket{\psi_1}$, and states with even larger gaps by optimizing over complex-valued states $\ket{\psi_1}$. Such examples with larger $\Delta(\rho_{AB})$ values are discussed in detail in Appendix~B, with the multipartite entanglement property of the states discussed in Appendix~C.

Let us point out that the non-zero gap $(1-R^{\rm out}(\rho_{AB}))$ in (\ref{RinRout}) implies a whole family of feasible states for cyclic EPR steering. To this end, we set $p=0.999$ in the one-parameter family of states $\rho(p)$ defined by (\ref{rho3q}). In this case, we have the two-qubit reduced state 
\begin{equation}
\rho_{AB}(p)=p\rho_{AB}+(1-p)\mathbb{I}_4/4. 
\label{rhoABp}
\end{equation}
Running the code of Ref.~\cite{nguyen} with $N=252$ for $\rho_{AB}(0.999)$ and $\rho_{BA}(0.999)$ gives the following bounds on the critical radius:
\begin{align}
R^{\rm out}(\rho_{AB}(0.999))&=0.99994147,\nonumber\\
R^{\rm in}(\rho_{BA}(0.999))&=1.0030571,
\label{RinRoutnoisy}
\end{align}
fulfilling both conditions in (\ref{R1R2}), which implies that $\rho_{AB}(0.999)$ is one-way steerable.

Let us now show that $\rho_{AB}(p)$ is one-way steerable not only for $p=1$ and $p=0.999$ but also for any $p\in\left[0.999,1\right]$. Indeed, according to the definition of the critical radius $R(\rho_{AB})$, the state $\rho_{AB}(R)=R\rho_{AB}+(1-R)\mathbb{I}/2\otimes \rho_B$ is on the border of unsteerable states. If we mix $(1-p)$ white noise to $\rho_{AB}(R)$ to get $p\rho_{AB}(R)+(1-p)\mathbb{I}_4/4$, the state is still unsteerable. Therefore the critical radius of the state $\rho_{AB}(p)$ in Eq.~(\ref{rhoABp}) is at least $R(\rho_{AB})$. 
Then our claim follows about one-way steerability of $\rho_{AB}(p)$ for $1\ge p\ge 0.999$, which in turn implies that the family of states~(\ref{rho3q}) with $\ket{\psi_1}$ in (\ref{optpsi1simp}) exhibits cyclic EPR steering in the range $1\ge p\ge 0.999$. 

\emph{Discussion.}---
We have shown the existence of three-party states arranged in a triangular configuration, where each two-party reduced state is steerable in one (e.g., clockwise) direction, but unsteerable in the other (e.g., anticlockwise) direction. That is by choosing any pair of systems out of the tripartite system belonging to parties $P$ and $P'$, steering can occur from party $P$ to party $P'$ if and only if $P'$ lies clockwise of $P$. This shows a peculiar directional feature
of EPR quantum correlations, which can neither appear in the phenomenon of entanglement nor in Bell nonlocality. To study this directional or handedness property of quantum correlations we have focused mainly on three-qubit states with projective measurements. However, we also discuss briefly the case of higher dimensional systems and the role of POVM measurements in Appendix~D. A couple of questions have been left open. One such question is whether our three-qubit cyclic steering result valid for projective measurements could also be extended to the most general form of POVM measurements. Second, is it true that cyclic steering for three-party systems always entails genuine tripartite entanglement? Third, it would also be interesting to generalize the construction of three-qubit translationally invariant states beyond three parties. From a more applied point of view, it would be interesting to find useful information applications of the directional feature of quantum correlations. Finally, a fully analytical proof of the three-qubit cyclic EPR steering phenomenon would be welcome.  

\emph{Acknowledgements.} We thank Wiesław Laskowski, Miguel Navascu\'es, G\'eza T\'oth and Jordi Tura for valuable discussions. EB and TV acknowledge the support of the EU (QuantERA eDICT) and the National Research, Development and Innovation Office NKFIH (No. 2019-2.1.7-ERA-NET-2020-00003).

\section{Appendix}

\emph{Appendix A: Measurement settings of Alice and Bob in setup 1.}---Let the following six unit vectors $\vec b_x$ ($x=1,\ldots,6$) correspond to Bob's measurement settings with outcome $b=0$ and their antipodal points $-\vec b_x$ to the outcome $b=1$:
\begin{align}
\vec b_1 &= [0, 1, \phi]/n, \quad\vec b_2 = [0, 1, -\phi]/n,\nonumber\\
\vec b_3 &= [1, \phi, 0]/n, \quad\vec b_4 = [1, -\phi, 0]/n,\nonumber\\
\vec b_5 &= [\phi, 0, 1]/n, \quad\vec b_6 = [\phi, 0, -1]/n,
\label{vecbx}
\end{align}
where $\phi= (1+\sqrt{5})/2$ is the golden ratio and $n=\sqrt{1+\phi^2}$ is a normalization constant. In particular, the $\pm \vec b_x$ vectors above form the vertices of the regular icosahedron. See also figure~\ref{fig_blochs} for the arrangement of the $\vec b_x$ vectors on the Bloch sphere (the respective endpoints are marked by small blue circles). 

On the other hand, Alice's Bloch vectors $\vec a_x$ corresponding to the outcome $a=0$ look as follows 
\begin{align}
\vec a_1 &= [0.1, 0.7664,   -0.6346],\nonumber\\
\vec a_2 &= [-0.1,  0.7664,    0.6346],\nonumber\\
\vec a_3 &= [0.4959,  0.8683,    0.0099],\nonumber\\ 
\vec a_4 &= [0.4959, -0.8683,    0.0099],\nonumber\\
\vec a_5 &= [0.9563,  0,        -0.2925],\nonumber\\
\vec a_6 &= [0,       0,         0].
\label{vecax}
\end{align}
In Fig.~\ref{fig_blochs}, the endpoints of the six $\vec a_x$ vectors are denoted by small red circles. 

\emph{Appendix B: Numerically obtained cyclic EPR steerable states in setup~2.}---We define the state $\ket{\psi_1}=\sum_{i,j,k=0}^{1}c_{ijk}\ket{ijk}$ according to Eq.~(\ref{psi1cijk}), which is used to define the state $\rho(p)$ in Eq.~(\ref{rho3q}). Here we set $p=1$, so we simply have
\begin{equation}
\rho_{ABC}\coloneqq \rho(1)=\frac{\sum_{i=1}^3 \ket{\psi_i}\bra{\psi_i}}{3},
\label{rho3peq1}
\end{equation}
where $\ket{\psi_2}$ and $\ket{\psi_3}$ are derived from  one- and two-qubit translations of $\ket{\psi_1}$ according to Eq.~(\ref{rightS}). Note that the reduced two-qubit states of $\rho_{ABC}$ above have the properties $\rho_{AB}=\rho_{BC}=\rho_{CA}$. Therefore, if we find that $\rho_{AB}$ is one-way steerable, then it also holds true for the other two marginals $\rho_{BC}$ and $\rho_{CA}$, which implies the existence of a tripartite cyclic EPR steerable state. We use the metric $\Delta(\rho_{AB})=R^{\rm in }(\rho_{BA})-R^{\rm out}(\rho_{AB})$ to the quality of the cyclic steerability property of the state~(\ref{rho3peq1}), where the values of the feasible critical radii are restricted to $R^{\rm in }(\rho_{BA})\ge 1$ and $R^{\rm out}(\rho_{AB})<1$. Then according to the conditions~(\ref{R1R2}), $\Delta(\rho_{AB})>0$ implies that $\rho_{AB}$ is one-way steerable for projective measurements and this in turn implies that $\rho_{ABC}$ is a cyclically steerable state. In what follows, we present three such states, denoted $\rho_{ABC}^{(i)}$, where each state is defined by $\ket{\psi_1^{(i)}}$ for $i=1,2,3$ in Eq.~(\ref{rho3peq1}).

Our first state $\rho_{ABC}^{(1)}$ corresponds to the state~(\ref{optpsi1simp}) in the main text. The coefficients $c_{ijk}$ of this state are also shown in the second column of Table~\ref{tab1}. Here we set $c_{111}=0$ and the other seven coefficients are real-valued. The corresponding reduced two-qubit state $\rho_{AB}^{(1)}$ (and its swapped state $\rho_{BA}^{(1)}$) has the following critical radius parameters 
\begin{align}
R^{\rm out}\left(\rho_{AB}^{(1)}\right)&=0.99822006,\nonumber\\
R^{\rm in}\left(\rho_{BA}^{(1)}\right)&=1.0000028,
\end{align}
resulting in $\Delta(\rho_{AB}^{(1)})=0.00178274$. By direct calculation, the negativity of this state is $E_N=0.0630$.

Our next state $\rho_{ABC}^{(2)}$ corresponds to $\ket{\psi_1^{(2)}}$ in Table~\ref{tab1} (third column). This is the best real-valued solution we have found, which has the following critical radius parameters: 
\begin{align}
R^{\rm out}\left(\rho_{AB}^{(2)}\right)&=0.99999963,\nonumber\\
R^{\rm in}\left(\rho_{BA}^{(2)}\right)&=1.00237752,
\end{align}
with $\Delta(\rho_{AB}^{(2)})=0.00237789$. On the other hand, the negativity of this state is $E_N=0.0679$.

Finally, the best complex-valued solution corresponding to the state  $\ket{\psi_1^{(3)}}$ is shown in the fourth column of Table~\ref{tab1}. This solution has the following critical radius parameters 
\begin{align}
R^{\rm out}\left(\rho_{AB}^{(3)}\right)&=0.99999898,\nonumber\\
R^{\rm in}\left(\rho_{BA}^{(3)}\right)&=1.0028066,
\end{align}
with $\Delta(\rho_{AB}^{(3)})=0.00280762$. The negativity of this two-qubit state is $E_N=0.0600$.

\begin{table}[t!]
	\begin{center}
		\begin{tabular}{|l|r|r|r|}
			\hline
			$c_{ijk}$ & $\ket{\psi_1^{(1)}}$ & $\ket{\psi_1^{(2)}}$ & $\ket{\psi_1^{(3)}}$\\
			\hline
			\hline
			$c_{000}$ &   $1$    &    $1$&      $1$\\
			\hline
			$c_{001}$ &   $-0.321193$    &    $-0.259910$&      $-0.252592-0.065698i$\\
			\hline
			$c_{010}$ &   $-0.477021$    &    $-0.591007$ &     $0.002913-0.000635i$\\
			\hline
			$c_{011}$ &   $0.045221$  &     $0.028007$ &     $0.025469+0.025479i$\\
			\hline
			$c_{100}$ &   $-0.718592$  &     $-0.798924$ &     $-0.120348-0.110323i$\\
			\hline
			$c_{101}$ &   $0.213715$  &     $0.206125$ &     $-0.103340-0.161335i$\\
			\hline
			$c_{110}$ &   $-0.0482$  &     $-0.079214$ &     $-0.044067-0.089806i$\\
			\hline
			$c_{111}$ &  $0$  &     $-0.000311$ &     $0.055929-0.044192i$\\
			\hline
		\end{tabular}
	\end{center}
	\caption{
		Table for the coefficients $c_{ijk}$ of $\ket{\psi_1^{(i)}}$ in Eq.~(\ref{psi1cijk}) for $i=1,2,3$.}
	\label{tab1}
\end{table}

\emph{Appendix~C: Entanglement properties of the numerically obtained cyclic EPR steerable states in scenario 2.}---Here we analyze the tripartite entanglement properties of the cyclic EPR-steerable states $\rho_{ABC}^{(i)}$, $i=1,2,3$ given in Appendix~B. In particular, we use the criteria of genuine tripartite entanglement (GTE) developed by G\"uhne and Seevinck~\cite{GS10} and the numerical method developed in Ref.~\cite{JMG11}. On the one hand, we have already given the negativity $E_N$ of the two-qubit reduced states of $\rho_{ABC}^{(i)}$ in Appendix~B. Since all $E_N(\rho_{AB}^{(i)})$ are strictly positive, it entails that all two-qubit marginals of $\rho_{ABC}^{(i)}$ are entangled. However, to the best of our knowledge, it is an open problem whether a given three-qubit state with all two-qubit marginals entangled implies that the three-qubit state is genuinely tripartite entangled (GTE). Indeed, multipartite entanglement properties can be very complex even for simple multipartite states, such as three-qubit states~\cite{horo_review,gt_review,szilard}.

Here we show that the complex-valued three-qubit state $\rho_{ABC}^{(3)}$ given in Appendix~B is genuinely tripartite entangled. To this end, we first define biseparable states. A three-party state $\rho_{ABC}$ can be written in biseparable form if and only if:
\begin{equation}
\rho_{ABC}=p\rho_{A|BC}+q\rho_{B|AC}+(1-p-q)\rho_{C|AB},
\label{bisep}
\end{equation}
where $p,q,(1-p-q)$ are non-negative numbers and $\rho_{A|BC}$ represents any tripartite biseparable state with respect to the cut $A|BC$ (and the other terms are defined similarly). If $\rho_{ABC}$ cannot be written in this form, we say that it is GTE.

We then show that the three qubit state $\rho_{ABC}^{(3)}$ is GTE by invoking the following sufficient criterion of GTE~\cite{GS10}:
\begin{equation}
|\rho_{1,8}|> \sqrt{\rho_{2,2}\rho_{7,7}}+\sqrt{\rho_{3,3}\rho_{6,6}}+\sqrt{\rho_{4,4}\rho_{5,5}},
\label{GScrit}
\end{equation}
where $\rho_{i,j}$ denotes the $(i,j)$ entry of a three-qubit density matrix $\rho$, where the standard product basis $\{\ket{000},\ket{001},\ldots,\ket{111}\}$ is assumed. Since for our particular state~$\rho_{ABC}^{(3)}$,  $|\rho_{1,8}|= 0.0621$ and the right-hand-side of (\ref{GScrit}) is $0.0588$, the criterion is readily satisfied. Therefore our state is GTE, as reported. 

Note also that we could not detect GTE for either $\rho_{ABC}^{(1)}$ or $\rho_{ABC}^{(2)}$ using the criteria of Ref.~\cite{GS10}. However, the numerical method of Ref.~\cite{JMG11} can detect GTE in these states as well. We then leave it as an open problem whether there exist three-party cyclically steerable states which are not GTE. In this respect, it is worth noting that for higher dimensional tripartite systems there exist biseparable states (i.~e. states that can be written in the form~(\ref{bisep})) in which all reduced bipartite systems are entangled. Such an example was presented in Ref.~\cite{persistency}. It remains to be proven, however, whether these states exhibit cyclic EPR steering.

\emph{Appendix D: Cyclic steering with POVMs and higher dimensional states.}---Here we show that if we do not restrict the local dimension to a qubit, we can strengthen our findings by generalizing cyclic EPR steering from projective measurements to general POVM's. To this end, we consider another construction, where the tripartite state is given by
\begin{equation}
\rho_{\tilde{A}\tilde{B}\tilde{C}}=\rho_{AB}\otimes\rho_{B'C}\otimes\rho_{C'A'},
\label{rho_abc}
\end{equation}
where the subsystems $\tilde{A}=AA'$, $\tilde{B}=BB'$, $\tilde{C}=CC'$ belong to Alice, Bob and Charlie, respectively. Let us further distribute the same states between the parties, that is, $\rho_{AB}=\rho_{B'C}=\rho_{C'A'}$. Furthermore, let $\rho_{AB}\in\CC^{d_A}\otimes\CC^{d_B}$ for some $d_A$ and $d_B$ be such that Alice can steer Bob's state using POVM measurements, but not the other way around (i.e. the state is one-way steerable for the most general POVMs). The existence of such a bipartite state with $d_A=2$ and $d_B=3$ was proven in Ref.~\cite{marco15}. More recently, the required dimension has been further reduced to $d_A=d_B=2$ in Ref.~\cite{bowles16}. In this case, the $(d\times d\times d)$-dimensional state $\rho_{\tilde{A}\tilde{B}\tilde{C}}$ with $d=d_A\times d_B=4$ exhibits the desired cyclic EPR steering property $A\to B$, $B\to C$, and $C\to A$. 

Indeed, in order for Alice to steer Bob (and also for Bob to steer Charlie and Charlie to steer Alice), the respective parties perform measurements on their entangled parts of subsystems.

On the other hand, the unsteerability property in the opposite direction follows from the fact that the composite state $\rho_{\tilde{A}\tilde{B}\tilde{C}}$ in Eq.~(\ref{rho_abc}) is product across the bipartitions $A|A'$, $B|B'$ and $C|C'$. Hence we have the reduced state $\rho_{\tilde{A}\tilde{B}}=\tr_{\tilde{C}}\rho_{\tilde{A}\tilde{B}\tilde{C}}=\rho_{AB}\otimes\rho_{A'}\otimes\rho_{B'}$, 
where $\rho_{A'}$ and $\rho_{B'}$ are the reduced states of $\rho_{CA'}$ and $\rho_{B'C}$, respectively. Due to convexity, if the state $\rho_{AB}$ admits a LHS model from Bob to Alice for POVM measurements, then the state $\rho_{\tilde{A}\tilde{B}}$ still admits a LHS model for POVM measurements. 

As mentioned above, using the one-way steerable state $\rho_{AB}\in\CC^2\otimes\CC^2$ from Ref.~\cite{bowles16}, we have a three-party cyclic EPR steerable state~(\ref{rho_abc}) for POVM measurements with dimension $(4\times 4\times 4)$. We state it as an open problem to find a cyclic EPR steerable state for POVM's in lower dimensions.

\end{document}